\documentstyle[sprocl]{article}

\input{psfig}

\def\be{\begin{equation}}
\def\ee{\end{equation}}
\def\bea{\begin{eqnarray}}
\def\eea{\end{eqnarray}}

\def\fun#1#2{\lower3.6pt\vbox{\baselineskip0pt\lineskip.9pt
  \ialign{$\mathsurround=0pt#1\hfil##\hfil$\crcr#2\crcr\sim\crcr}}}

\def\etal{{\it et. al.}}

\begin{document}

\title{Adiabatic CDM Models and the Competition}

\author{Lloyd Knox}

\address{Department of
Astronomy and Astrophysics, The University of Chicago\\5640 So. Ellis Avenue,
Chicago, IL~~60637-1433, USA}
\vskip 0.05in
\address{E-mail: knox@flight.uchicago.edu}


\maketitle\abstracts{The inflation-inspired flat, cold
dark matter-dominated models of structure formation with
adiabatic, nearly scale-invariant initial conditions agree very well
with current CMB anisotropy data.  The success of these models
is highlighted by the failure of alternatives; we argue
that there are no longer any viable competitors (with the
exception of models with more complicated matter content 
which are still flat and which still require inflation).  
CMB data will soon be of sufficient quality that, if one {\it assumes}
inflation, one can detect a non-zero cosmological constant by
combining a determination of the peak location with 
Hubble constant measurements.
}

\section{Introduction}

The aim of this paper is to demonstrate the success of
inflation-inspired models of structure formation.  The CMB data are
pointing us towards models in which the mean spatial curvature is
zero, and in which the ``initial'' perturbations were adiabatic and
nearly scale-invariant.  These three properties are all predictions of
the simplest models of inflation.  We will discuss them below and how
they influence the properties of the CMB.  Since we are never able to
prove a model to be true, just that it is more probable than other
models, much of the demonstration of the success of inflation-inspired
models is a discussion of what goes wrong with other ones.

We begin with a very quick review\cite{swrev} of the basics
and then move on to a brief description of current data.
The subsequent discussion of adiabatic models explains what adiabatic,
flat and nearly scale-invariant mean and how these properties
influence the CMB power spectrum.  With this discussion complete we
are then ready to see how isocurvature and defect models 
differ, and that they do so in ways that conflict with the data.  
Finally, the strong constraint on the peak
location given all data, prompts a discussion about what we can learn
from the peak location besides the geometry.

\section{Preliminaries}

At sufficiently early times, a thermal distribution of photons kept
all the atoms in the Universe ionized.  Because of the strength of the
Thomson cross section and the large number density of electrons, the
photons were tightly coupled to the electrons (and through them to the
nuclei) and therefore these components could be treated as a single
fluid called the photon-baryon fluid.  As the photon temperature
cooled (due to the expansion of the Universe) below one Rydberg
(actually well below one Rydberg due to the enormous photon-to-baryon
ratio), the electrons combined with the nuclei thereby decoupling the
photons from the baryons.  The Universe became transparent to the
photons that are now the CMB.  Thus when we look at the CMB, we are
seeing the Universe as it was at the time of decoupling---also
referred to as ``last-scattering''.

The temperature of the CMB is the same in all directions, to 1 part in
100,000.  The most interesting statistical property of these tiny
fluctuations is the angular power spectrum, $C_l$, which tells us how
much fluctuation power there is at different angular scales, or
multipole moments $l$ (where $l \sim \pi/\theta$).  Because the
departures from isotropy are so small, linear perturbation theory is
an excellent approximation and the angular power spectrum can be
calculated for a given model with very high precision.  Thus the CMB
offers a very clean probe of cosmology---one where the basic physics
is much better understood than is the case for galaxies or even
clusters of galaxies.

Throughout, $\Omega_i$ is the mean density
of component $i$, $\bar \rho_i$, in units of the critical density
which divides negatively and positively curved models.
Note that $\Omega \equiv \sum_i \Omega_i = 1$ corresponds to the
case of zero mean spatial curvature.

\section{The Data}

The last year of the 1000's was a very exciting one for those
interested in measurements of the angular power spectrum.  
New results came from MSAM\cite{wilson99}, PythonV\cite{coble99a,coble99b}, 
CAT\cite{bak99}, MAT\cite{mat97,mat98,tocoexplained}, 
IAC\cite{dicker99}, Viper\cite{peterson}, and BOOM/NA\cite{mauskopf}, all of
which have bearing on the properties of the peak.  
These data make a convincing case
that we have indeed observed a peak---which not only rises towards
$l=200$ (as we have known for several years\cite{risedetect}) but also falls
dramatically towards $l=400$.  Figure 1 shows the results from 1999
plus, in background shading, a fit\cite{BJKII} 
of the power in 14 bands of $l$ to
all the data.  Many of the bands are at
low enough $l$ that they cannot be discerned on a linear x-axis plot.
The $\Omega=1$ model in the figure 
has a mean density of non-relativistic matter, 
$\Omega_m=0.31$, a cosmological constant density of $\Omega_\Lambda=0.69$,
a baryon density of $\Omega_b=0.019h^{-2}$ \protect\cite{BNTT99}, a Hubble
constant of $H_0 = 100 h\ {\rm km/sec/Mpc}$ with $h=0.65$, 
an optical depth to reionization
of $\tau = 0.17$ and a power spectrum power-law index of $n=1.12$, where
$n=1$ is scale invariant.

\begin{figure}
\label{fig:data}
{\psfig{figure=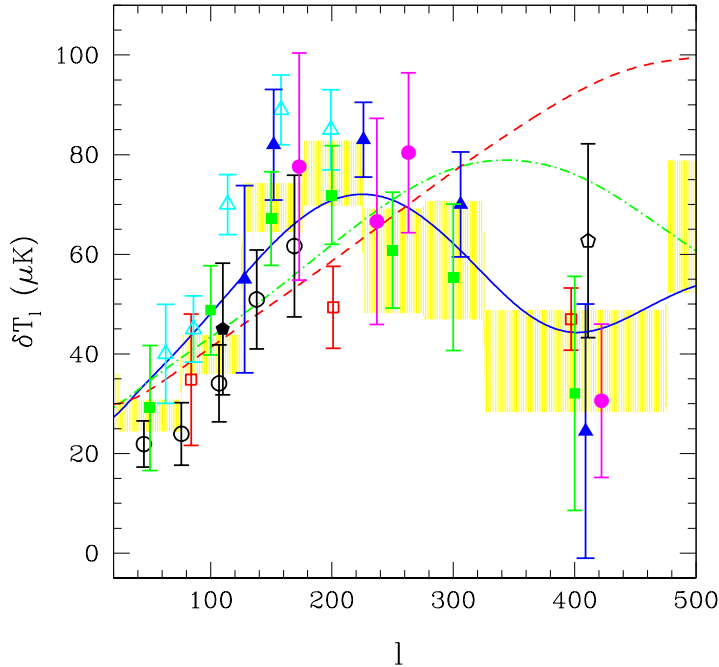,width=4in}}
\caption{Bandpowers from TOCO97 (cyan open triangles), 
TOCO98 (blue filled triangles), BOOM/NA (green filled squares),
MSAM (red open squares), CAT (black open pentagon), IAC (black 
filled pentagon), PyV (black open circles) and Viper
(green filled circles). The y-axis is 
$\delta T_l \equiv \sqrt{l(l+1)C_l/(2\pi)}$ where $C_l$ is the
angular power spectrum. The models
are (from peaking at left to peaking at right) the best fit models
of \protect\cite{dodknox99} for $\Omega=1$, $\Omega=0.4$ and $\Omega=0.2$.  
Calibration errors are not shown.  
}
\end{figure}

Knox and Page \cite{KP00} have recently characterized the peak with
fits of phenomenological models to the data.  They find the peak to
be localized by TOCO and BOOM/NA at $175 < l_{\rm peak} < 243$ and
$151 < l_{\rm peak} < 259$ respectively (both ranges 95\% confidence).
This location is also indicated by combining the PythonV and Viper
data, as can be seen in Fig. 1 and a significant bound can also be
derived by combining all other data, prior to these four data sets.  
In sum, a peak near $l \sim 200$, is robustly detected.  
Combining all the data, we can also constraint its full-width at 
half-maximum to be between 180 and 250 at 95\% confidence.  

\section{Physical Models}

In this section we describe three classes of physical models
(adiabatic CDM, topological defects and isocurvature dark matter) and
their predictions for the angular power spectrum\cite{whatweknow}.  

\subsection{Adiabatic CDM}

The simplest models of inflation lead to a post-reheat Universe with
critical density (to exponential precision) and adiabatic, nearly
scale-invariant fluctuations\cite{KT}.  Although inflation does not require
cold dark matter, the prediction of critical density, combined with
upper limits on the mean baryonic density, push one in that direction.
Within the last few years, the observations have developed to strongly
prefer the gap between $\Omega_b$ $(b \equiv {\rm baryons})$
and unity to be filled with not just cold dark matter, 
but a sizeable helping of dark energy too, e.g., 
\cite{dodknox99,bahcall,mturner,SNe}.
Let's examine more precisely each of these three predictions.

\subsubsection{Adiabatic}

Adiabatic (or, equivalently, isentropic) means that there are no
spatial fluctuations in the total entropy {\it per particle of each type}.  
That is, $\delta
(s/n_i) = 0$ for all species $i$.  From this, we see that $\delta
n_i/n_i = \delta s/s$ and therefore all species have the same
fractional fluctuation in their number densities.  For example, where
there are more dark matter particles there are more photons, etc.  A
general perturbation\cite{generalperturb} is a
linear combination of isocurvature and adiabatic modes.  Isocurvature
perturbations are arranged so that $\delta \rho = \sum_i \delta \rho_i = 0$.

The evolution of a single Fourier mode is initially a competition
between the pressure of the baryon-photon fluid trying to decrease
density contrasts, and gravity trying to enhance them.
For adiabatic modes the gravitational term is initially dominant, 
increasing the amplitude of the mode, until the restorative
force of the pressure gradients pushes it back. Despite the
initial growth in the density contrast, the potential decays.
This
is because the photon pressure prevents the growth from happening
quickly enough to counteract the effects of the expansion.  It is this
decay of the potential that leads to excitation of a cosine mode
(after the initial transient) for the acoustic oscillation.  For
isocurvature modes, the potential is initially zero, until there is
sufficient time for pressure gradients to evolve into density
gradients.  The initially growing potential excites a sine mode
\cite{husug,HW}.

All adiabatic modes of given wavenumber, $k=\sqrt{k_x^2+k_y^2+k_z^2}$,
although they have different {\it spatial} phases, will all have the
same {\it temporal} phase, because they all start off with the same
relationship between the dark matter and photons.  In other words,
although spatially incoherent, they are temporally coherent.  This
coherence is essential to the familiar Doppler peak structure of the
CMB power spectrum \cite{coherence}.  
Figure 2 illustrates the point by showing the
spatial dependence of three different modes with varying wave numbers
at five different stages of their evolution.  

\begin{figure}
\centerline{\psfig{figure=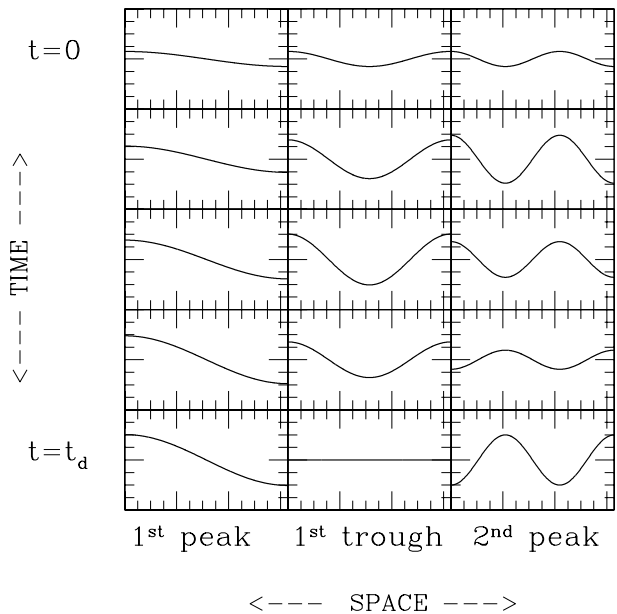,width=3.5in}}
\caption{Spatial dependence of the photon-baryon fluid temperature
for three different modes at five different times.  Shown is the evolution
from very early when the wavelengths are much larger than the Hubble
radius and causal processes have not had time to change
the ``initial conditions'' as left by inflation (top panels) to the
time of decoupling between matter and radiation (bottom panels).  
The longest wavelength mode is shown on the left.  
It reaches its
first maximum just at the time of decoupling.  The peak-to-trough
distance, as seen by us today, subtends about a degree on the
last-scattering surface.  The fact that this mode (and all modes 
with the same wavelength) reach an extremum just at the time
of decoupling is the reason for the first peak in the
CMB power spectrum at $l \sim 200$.  Contrast this behavior 
with that of a mode with half the wavelength
(shown in the five middle panels) which therefore oscillates in
time twice as fast and hits a null at the time of decoupling.
Modes like this one have a peak-to-trough distance of about half
a degree and are responsible for the first trough at $l=400$.
Finally, the right-most panels are for the 
even faster and shorter modes responsible for the
second peak.
}
\end{figure}

\subsubsection{Flat}

Inflation generically produces flat Universes---ones where the mean
spatial curvature is exponentially close to zero (e.g., $e^{-100}$ is
a particularly large residual curvature).  The CMB is sensitive to
curvature because the translation of a linear distance on the last-scattering
surface to an angular extent on the sky depends on it.  
One can see this by noting that
in a negatively curved space the surface area of a sphere of radius
$r$ is larger than $4\pi r^2$ and therefore objects at fixed
coordinate distance of fixed size appear smaller than they would in
the case of zero curvature (the larger-than-$4\pi r^2$ sphere must be
squeezed into a local $4\pi$ steradians).  This geometrical effect
shifts the CMB power spectrum peak locations by a factor of
$\Omega^{-1/2}$.  Other parameters also affect the peak locations by
altering the coordinate distance to the last-scattering surface and
the size of features there; these are subdominant effects which will
be discussed later.

\subsubsection{Nearly Scale Invariant}

The power spectrum of fluctuations produced by the simplest models of
inflation is
well-described by a power law, $P(k) \propto k^n$, with $n$ near
unity.  The case $n=1$ is called scale-invariant because the
dimensionless quantity $k^3 P(k)$ is the same for all modes when the
comparison is done at ``horizon crossing'' (when the mode wavelength
becomes smaller than the Hubble radius)\cite{KT}.

\subsection{Topological Defects}

The usual scenario for topological defects is that a phase transition
in an initially homogeneous Universe gives rise to a scalar field with
a spatially-varying stress-energy tensor.  In most such scenarios, the
scalar field configuration evolves into a network of regions which are
topologically incapable of relaxing to the true ground state.
Causality implies that these models have isocurvature initial
conditions.

In defect models, the temporal coherence is lost due to continual
sourcing of new perturbations by the non-linearly evolving scalar
field.  This generically leads to one very broad peak with a maximum
near $l=400-500$\cite{coherence}.  Thus the drop in power from $l=200$
to $l=400$ is a very challenging feature of the data. 
One can get lower power at $l=400$ than at $l=200$ with
modifications to the ionization history\cite{WBA}, but even for
these models the drop probably is not fast enough.  See A. Albrecht's
contribution to these proceedings.

\subsection{Isocurvature Dark Matter Models}

Note that isocurvature is more general than adiabatic.  Given numerous
components, there are a number of different ways of maintaining the
isocurvature condition, $\sum_i \delta \rho_i = 0$.  In what follows we
will assume that the isocurvature condition is maintained by the dark
matter compensating everything else.

Isocurvature models have at least two strikes against them.  First,
scale-invariant models produce far too much fluctuation power on large
angular scales, when normalized to galaxy fluctuations at smaller
scales as has been known for over a decade \cite{EB86}.  
One might
hope to save isocurvature models by tilting them far from
scale invariant, but this fix cannot simultaneously get galaxy scales,
COBE-scales and Doppler peak scales right.

The second strike has to do with the location of the acoustic peaks.
As mentioned above, the isocurvature oscillations are 90 degrees out
of phase with the adiabatic ones.  The peaks get shifted to {\it
  higher} $l$ ($l = 350$ to $400$ for the first peak) \cite{HW}.  Geometrical
effects could shift it back, to make it agree with the data, but this
would require $\Omega > 2$ which is inconsistent with a number of 
observations\cite{whatweknow}.  

There are scenarios with initially isocurvature conditions that can
produce CMB power spectra that look much like those in the adiabatic
case.  This can be done by adding to adiabatic fluctuations, another
component which maintains the isocurvature condition and then by
giving this extra component a non-trivial stress history
\cite{postmodernisocurv}.  These alternatives will be interesting to
pursue further if improvements to the data cause troubles for the
currently successful adiabatic models.  Even these alternatives are
flat models that require some mechanism, such as inflation, for
creating the super-horizon correlations in their initial conditions.

Turok has shown that even super-horizon correlations are not a necessary
condition for CMB power spectra that mimic those of inflation\cite{turok}.
Although no specific model is constructed, this work demonstrates that
causality alone does not preclude one from getting inflation-like
power spectra without inflation.  For a discussion of the physical plausibility
of models that could do this, see \cite{HSW97}.

\section{Peak Location and $\Omega_\Lambda$}

If we {\it assume} 
flatness, adiabaticity and near scale-invariance we can 
then determine $\Omega_\Lambda$ from the location of the first peak
\cite{marc}.
With these assumptions, the peak position just depends on the
coordinate distance to the last-scattering surface divided by the
sound horizon at last-scattering.  How this ratio depends on $\Omega_\Lambda$
depends on what else we hold fixed.  If we have high-precision
CMB data over several peaks then $w_b \equiv \Omega_b h^2$ and
$w_c \equiv \Omega_c h^2$ would be good things to keep fixed, since
those are what affect the acoustic peak morphology\cite{EB99}.  
However, without
such high precision data, $w_b$ and $H_0$ are good things to fix because
we know these fairly well from other measurements.  With $w_b$ and $H_0$
fixed, increasing $\Omega_\Lambda$ increases the sound horizon (because
$w_c$ must decrease) but increases the coordinate distance to the
last-scattering surface by more and the peak moves out to higher $l$.  With
$w_b$ and $w_c$ fixed, the sound-horizon stays the same but $H_0$ increases
and the coordinate distance to the last-scattering surface drops: the
peak shifts to lower $l$.

\begin{figure}
\label{fig:lambdaVlpeak}
\centerline{\psfig{figure=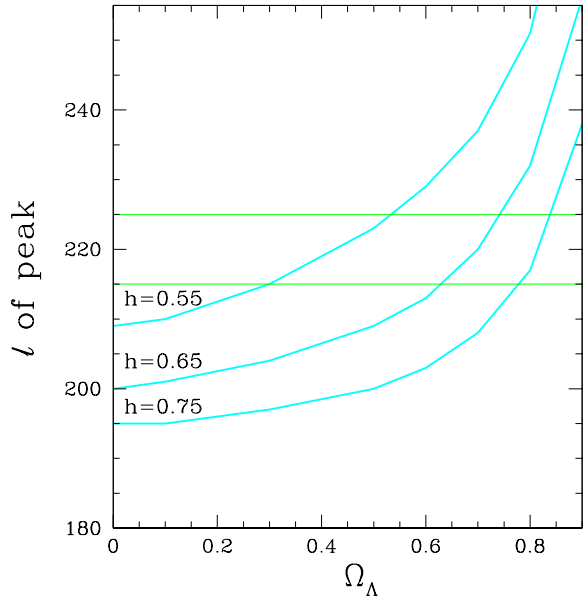,width=3.5in}}
\caption{We find $l_{\rm peak}$ as we vary $\Omega_\Lambda$
for three values of the Hubble constant.  We always fix
$\Omega_b h^2 = 0.019$, no reionization, no gravity waves, no tilt.
The horizontal lines show 1$\sigma$ constraint on $l_{\rm peak}$ of
$220 \pm 5$ that
should be possible from Boom98 data, just using $100< l < 300$.  The
corresponding bound for MAP will be about $\pm 1$.
}
\end{figure}

If we take all the data, the peak
location is $l_{\rm peak} =229\pm 9$\cite{KP00}.  
Thus, if we assume $h=0.65$ then
this is evidence for a positive cosmological constant.  It is weak
evidence because inclusion of possible systematic errors would 
probably widen the peak bound significantly.    
However, for a sample-variance dominated measurement of the first
peak (specifically, $C_l$ from
$l=100$ to $l=300$) derived from observations of 1000 square
degrees of sky (comparable to the Antarctic Boomerang coverage)
one can determine the peak to be at, e.g.,  $l_c = 220 \pm 5$.   
One can see from Fig.~\ref{fig:lambdaVlpeak} that we may soon have
a strong determination of non-zero $\Omega_\Lambda$ based
solely on Hubble constant and CMB measurements.  Note that
this determination 
will not suffer from calibration uncertainty.

\section{Conclusion}
The peak has been observed by two
different instruments, and can be inferred from an independent
compilation of other data sets.  The properties of this peak are
consistent with those of the first peak in the inflation-inspired
adiabatic CDM models, and inconsistent with competing models, with the
possible exception of the more complicated isocurvature models
mentioned above.  It is perhaps instructive that where the
confrontation between theory and observation can be done with a
minimum of theoretical uncertainty, the adiabatic CDM models have been
highly successful.

\section*{Acknowledgments}
I thank A. Albrecht, L. Page and J. Ruhl for useful conversations and
D. Eisenstein for comments on the manuscript.
I used CMBAST\cite{cmbfast} and am supported by the DoE, NASA grant NAG5-7986,
and NSF grant OPP-8920223.

\end{document}